# Optomechanical photon shuttling between photonic cavities

Huan Li and Mo Li[1*]

[1]*Department of Electrical and Computer Engineering, University of Minnesota, Minneapolis, MN 55455, USA*

**Mechanical motion of photonic devices driven by optical forces provides a profound means of coupling between optical fields. The current focus of these optomechanical effects has been on cavity optomechanics systems in which co-localized optical and mechanical modes interact strongly to enable wave-mixing between photons and phonons and backaction cooling of mechanical modes. Alternatively, extended mechanical modes can also induce strong nonlocal effects on propagating optical fields or multiple localized optical modes at distances. Here, we demonstrate a novel multi-cavity optomechanical device: a "photon see-saw", in which torsional optomechanical motion can shuttle photons between two photonic crystal nanocavities. The resonance frequencies of the two cavities, one on each side of the see-saw, are modulated anti-symmetrically by the device's rotation. Pumping photons into one cavity excites optomechanical self-oscillation which strongly modulates the inter-cavity coupling and shuttles photons to the other empty cavity during every oscillation cycle in a well regulated fashion.**

Photonic cavities are indispensable components in integrated optical systems as buffers or registers to trap and store photons for processing both classical and quantum information[1-4]. To achieve advanced optical functions, it is often necessary to couple multiple cavities so that photons can be exchanged between them to achieve signal processing and computational operation[5-8]. Two cavities in proximity couple evanescently so that photons in one cavity can spontaneously couple to the other cavity with a coupling rate that is determined by the spatial overlap of the optical modes of both cavities. From the perspective of optical signal processing, however, it is desirable to couple cavities that are far separated so that each cavity can be individually controlled and the exchange of photons between them can be regulated. For example, two cavities can be coupled to a common bus waveguide to achieve strong enough coupling to cause photon number Rabi oscillation between the two; modulation of inter-cavity coupling can be achieved by detuning one of the cavities via optical injection of carriers[9].

* Corresponding author: moli@umn.edu

Recently, optical forces in nano-optomechanical systems have been exploited as a new means of controlling the coupling between optical modes in various nanophotonic systems[10-19]. In particular, cavity optomechanics explores the interactions between the cavity optical modes and the localized mechanical modes, such as the vibration of a mirror in a Fabry-Perot cavity[20-25] and the bulk and surface acoustic modes in whispering gallery resonators and photonic crystal cavities[26-29]. In these systems, there is only one optical cavity although multiple optical modes of the cavity have been utilized to induce the optomechanical effects. Meanwhile, multiple mechanical modes can also be synchronized through their optomechanical coupling with one cavity mode[30,31]. More recently, coupling between two Fabry-Perot cavities mediated by the mechanical motion of a nanomembrane in-between was theoretically studied[32]. In most of the systems mentioned above, however, the involved mechanical modes are highly localized compared to the spatial extent of the optical modes: the mechanical motion only perturbs a portion of the optical mode of the cavities. If instead of localized mechanical modes, an extended mechanical mode that simultaneously interacts with multiple cavities can be exploited, intriguing cavity optomechanical effects can emerge. In this work, we demonstrate a novel torsional multi-cavity optomechanical system, which we name a "photon see-saw", as depicted in Fig. 1a. The photon see-saw consists of a nanobeam inscribed with two 1D photonic crystal nanocavities, one on each side. Optical force generated by photons inside the nanocavities excites rotational motion of the nanobeam which anti-symmetrically modulates the frequencies of both nanocavities. With this device, we demonstrate optomechanical modulation of the coupling between the two nanocavities and the photon shuttling effect in the side-band unresolved regime. Our work focuses on well regulated transfer of photons in small amount (potentially and ultimately at single photon level), and represents an important first step toward the side-band resolved regime, where a wealth of intriguing photon dynamics can be explored[32].

Instead of the commonly used dielectric resonant modes, the two 1D photonic crystal nanocavities are identically designed to resonate in the fundamental air-mode (Fig. 1d) (effective mode index $n_{eff}$ = 1.71) which enables stronger coupling with the substrate and consequently higher optomechanical coupling coefficient[33,34]. (The design of the photonic crystal nanocavity is discussed in the Supplementary Information (SI).) On the mechanical design, the nanobeam is suspended on two axial spokes which are anchored to the substrate. The nanobeam can oscillate as a torsional resonator driven by the optical force generated by photons in either of the two cavities. Very unique in this torsional cavity optomechanics device is that the rotation of the nanobeam as an extended mechanical mode couples the two nanocavities dispersively in an anti-symmetric way: when the nanobeam rotates such that one side of it approaches the substrate, the cavity on the same side will experience a resonance red-shift due to increased coupling with the substrate while the resonance of the other cavity on the opposite side will be blue-shifted due to decreased coupling with the substrate. This anti-symmetric optomechanical effect is manifested in Fig. 1e where the resonance frequencies of the left and right cavities are simulated with varying rotation angle of the nanobeam. The linear and angular optomechanical coupling coefficients of the cavity ($g_{OM} = d\omega/dz$ and $g_{OM}^A = d\omega/d\theta = g_{OM} \cdot l$, where $l$ is the distance from the center of the cavity to the rotation axis of the nanobeam) are calculated to be $(2\pi) \cdot 2.13$ GHz/nm and $(2\pi) \cdot 24.5$ GHz/mrad, respectively, at zero rotation angle. Note that because the torsional mode is very compliant (effective spring constant $k = 0.11$ N/m), the cavity frequency shift induced by a single photon given by $\delta\omega_c = \hbar g_{OM}^2/k = 2g_0^2/\Omega_m = (2\pi) \cdot 27$ kHz is among the highest in cavity optomechanics systems demonstrated so far[19] and thus the design is favorable to achieving efficient optomechanical tuning of the cavity modes.

The photon see-saw device was fabricated on a standard silicon on insulator (SOI) substrate with a 220 nm top silicon layer and a 3 μm buried oxide (BOX) layer. After patterning the device with electron beam lithography and dry etch, the nanobeam was selectively released from the substrate by wet etching the BOX layer using hydrofluoric acid. Fig. 1b shows the optical microscope image after the nanobeam was released, whereas Fig. 1c shows the scanning electron microscope (SEM) image before the releasing to prevent the collapse of the nanobeam due to electrostatic charging induced in the SEM. Each of the two cavities is coupled with a waveguide, allowing them to be characterized independently. To avoid over-loading the cavity, the gap between the photonic crystal cavities and the waveguides is designed to be 500 nm, much larger than the 250 nm gap between the cavities and the substrate. Therefore, the optomechanical coupling between the cavity and the waveguide is negligible compared to that between the cavity and the substrate, which dominates the optomechanical effects in the device. Consequently, the out-of-plane motion driven by the substrate coupled optical forces dominates the in-plane motion induced by the waveguide coupled optical forces. In the transmission spectra measured from the two waveguides (Fig. 1f and g), the resonance modes of the two nominally identical cavities are shown to be very close in wavelength (left: $\lambda_{L0} = 1541.574$ nm; right: $\lambda_{R0} = 1541.219$ nm) with a detuning of $(2\pi) \cdot 44.8$ GHz. They also have similar quality factors (waveguide loaded: $Q_L \sim 1.0 \times 10^4$; intrinsic: $Q_i \sim 1.6 \times 10^4$) which are significantly lower than the simulated value ($Q \sim 10^6$) due to fabrication non-idealities. The two cavities are separated by a long distance of 23 μm, equivalent to about 50 optical wavelengths in silicon. Therefore, the coupling rate between the two cavities is very low ($\kappa \sim (2\pi) \cdot 0.72$ GHz) and thus the resonance mode of the right cavity is not observed in the spectrum measured from the left cavity and vice versa.

To investigate the mechanical and optomechanical properties of the device, we first employed an impulse measurement technique. One laser tuned to the resonance of the left cavity and modulated with a pulse generator was sent to the left cavity and used as a pulsed pump with effective pulse width of ~10 ns and peak power level of ~4 mW. Another continuous wave (CW) probe laser was input to the right cavity with varying detuning $\delta_b$, relative to the right cavity's initial resonance, and a low fixed power level of ~30 nW; its transmission through the device was monitored in time-domain so that the mechanical motion of the torsional nanobeam was readout with such a slope detection scheme. In Fig. 2a, the measured impulse responses when the device was in vacuum ($1 \times 10^{-4}$ Torr) and the probe laser was red and blue detuned ($\delta_b = \pm 0.8$) are displayed, showing the typical ring-down response of a mechanical resonator. Fig. 2b shows the Fourier transform spectra of the time-domain signals, revealing two dominant resonance modes at $\Omega_m = 2\pi \cdot 441$ kHz and $2\pi \cdot 514$ kHz which correspond to the fundamental torsional and flapping modes (insets, Fig.2b), respectively. As expected from the design, no in-plane mechanical modes can be observed in the spectra. In Fig. 2a, it is obvious that the ring-down time when the probe laser is blue detuned ($\delta_b > 0$) is longer than the ring-down time when the probe laser is red detuned ($\delta_b < 0$), indicating modified mechanical damping coefficients of both mechanical modes. This is due to the well-known backaction cooling and amplification effects induced by the detuned probe laser, which have been extensively investigated in various cavity optomechanics systems. The intrinsic mechanical quality factors, determined independently from thermomechanical noise calibration, are $1.66\times10^4$ and $1.68\times10^4$ for the torsional and flapping modes, respectively. Although the impulse response measurement in vacuum yields important information about the mechanical properties of the device, some interesting optomechanical effects are submerged by the fast oscillation. We next conducted the same impulse response measurement at atmospheric pressure utilizing air damping to "slow down" the mechanical motion as shown in Fig. 2c. Immediately after the pump pulse entered the left cavity ($t = 0$), a positive (negative) response was observed in the output of the probe laser red (blue) detuned from the right cavity resonance. As illustrated in Fig. 2d the sign of the response indicates that an instantaneous counter-clockwise rotation of the nanobeam and a consequent blue-shift of the right cavity's resonance were induced by the pulse of optical force generated at the left cavity by the pump pulse. Thus, this time-domain result reveals the interesting photon see-saw effect: the photons in the left cavity outweigh the photons in the right cavity, tilt the torsional nanobeam and detune the right cavity. The see-saw response reached a peak value at 1.2 μs and started to decrease as the nanobeam recoiled back. At 1.9 μs the response changed sign and peaked with much higher amplitude at 3.2 μs. This is because the pump pulse in the left cavity generated a heat pulse that propagated across the nanobeam to the right cavity and induced a thermo-optical red-shift of its resonance. The time scale of this event agrees well with the simulated thermal response of the device (see SI). From this impulse response measurement, the anti-symmetric nature of the optomechanical coupling between the nanobeam's torsional motion and the optical modes of the two cavities, i.e. the photon see-saw effect, is clearly revealed.

When the pump laser was changed to CW mode, significant cavity optomechanical backaction effects on the torsional mode of the nanobeam resonator were induced. Here we focus on the amplification effect (cooling effect is discussed in the SI), which can excite optomechanical self-oscillation with lasing-like characteristics[35]. Because of the very high mechanical quality factor of the torsional mode and the strong optomechanical coupling of the cavities, optomechanical self-oscillation can be triggered at a very low laser power level. During the experiment, the pump laser power sent to the left cavity was fixed at 3.4 μW to excite stable oscillation. The probe laser coupled with the right cavity was set at a much lower power level of 2.3 nW so its backaction effect was negligible compared to the pump. Fig. 3a shows the time-domain traces of the transmitted probe signal when stable oscillation was excited and the initial probe laser detuning $\delta_b$ was varied from positive to negative values. When the nanobeam oscillates in see-saw motion, the resonance frequencies of both cavities swing back and forth together but 180° out-of-phase with each other. The time-domain trace of the probe laser transmission can provide a real-time monitor of the right cavity's resonance: minimum when its resonance swung closest to and maximum when furthest away from the probe laser wavelength. This dynamics is illustrated in the right column in Fig. 3a. By varying the initial probe laser detuning and fitting the measured traces with the theoretical model (Fig. 3a, solid lines, see SI for details), stroboscopic snapshots of the cavity resonance frequency and line shape at any moment during the oscillation cycle can be constructed. The results are shown in Fig. 3b where the resonances of the left and the right cavities when the nanobeam rotates to the most counter-clockwise, the neutral and the most clockwise positions are plotted. From these results, it is clear that even with a modest pump power, the nanobeam can be excited into see-saw oscillation with amplitude high enough such that the resonances of the two cavities can cross over each other during each oscillation cycle. In Fig. 3c, the relative frequency shift of both cavities determined from above experiment is plotted versus the rotation angle of the nanobeam. Unlike the theoretical results (Fig. 1e), the actual optomechanical characteristics of the two cavities are not perfectly anti-symmetric despite of the identical design, which we attribute to the variance of the surface profile of the etched substrate under the suspended cavities. The result in Fig. 3c shows that the resonances of the two cavities align when the nanobeam is rotated to an angle of −0.7 mrad at which strong inter-cavity coupling can occur.

We demonstrate in the following that due to the unique photon see-saw effect and its strong modulation of the cavity resonances, photons can be optomechanically shuttled from the filled left cavity to the empty right cavity when self-oscillation with sufficient amplitude is excited. Such an optomechanical photon shuttling effect was previously predicted in a membrane-inside Fabry-Perot cavity optomechanical system[32], which has similar anti-symmetric optomechanical coupling between the two cavity modes. When the probe laser was removed and only the pump laser was on, the right cavity remained mostly empty because the inter-cavity coupling rate ($\kappa = (2\pi) \cdot 0.72$ GHz) was significantly lower than the cavity photon decay rate ($\gamma_{L,R} \sim (2\pi) \cdot 19$ GHz) and the two cavities were initially detuned by $(2\pi) \cdot 44.8$ GHz. The intra-cavity photon number $n_R$ of the right cavity is given by:

$$n_R = \frac{|a_R|^2}{\hbar\omega_p} = \frac{(\kappa\tau_R\tau_L)^2/\tau_{Le}}{(\delta_R^2+1)(\delta_L^2+1)}\left(\frac{P_{in}}{\hbar\omega_p}\right),$$

where $\delta_R$ ($\delta_L$) and $\tau_R = 2/\gamma_R$ ($\tau_L = 2/\gamma_L$) are normalized laser detuning and field decay time of the right (left) cavity, $P_{in}$ is the input pump laser power to the left cavity (see SI). Fig. 4a shows the time-domain trace of $n_R(t)$ and the transmission out of the right cavity when $P_{in}$ was reduced to the threshold level (~0.135 µW) of self-oscillation. It clearly shows that once per oscillation cycle a pulse of photons is shuttled from the left cavity to fill the right cavity. In between the pulses, the right cavity remains empty ($n_R \sim 0$) because the photon decay rate is much higher than the oscillation frequency, i.e. side-band unresolved. At the same time the photon number in the left cavity $n_L$ also oscillates because its detuning relative to the fixed pump laser frequency changes during the oscillation cycle (Fig. 4b). Thus, the photon number in the right cavity reaches a peak value every time the three frequencies of the left cavity resonance, the right cavity resonance and the pump laser align. As explained in Fig. 3, when the pump laser power increases, the oscillation amplitude grows and the resonance frequencies of the two cavities cross over during each oscillation cycle. As a result, two peaks per cycle start to appear in the time-domain trace of $n_R(t)$ in Fig. 4c when the pump laser power is gradually increased from 0.135 µW to 6.76 µW. Fig. 4d shows that the integrated number of shuttled photons $n_{tr}$ during an oscillation cycle increases with the pump laser power. At the threshold pump power level of 0.135 µW, ~1000 photons are shuttled to the right cavity during each cycle. The photon shuttling effect between two optomechanically coupled cavities can be modelled with the temporal coupled mode theory (described in SI). In Fig. 4e, the calculated number of photons in the right cavity $n_R$ (normalized) is plotted versus $\delta_R$ and $\delta_L$. In Fig. 4f, the normalized experimental results from the time-domain traces, which were obtained when the pump laser power was fixed at 6.76 µW and its detuning was varied, are plotted as trajectories with the two cavities' detuning determined from the stroboscopic method used in Fig. 3 (explained in the SI). Close agreement between theoretical model and experimental results can be observed. In Fig. 4e, the trajectory that the system undergoes in the parameter space when it is oscillating at the threshold level (i.e. the trace in Fig. 4a) is overlaid on the theoretical plot, illustrating the dynamics of photon shuttling process.

The demonstrated modulation of inter-cavity coupling and photon shuttling between two photonic cavities are mediated only by the optomechanical self-oscillation without any external modulation. It is thus very unique to cavity optomechanics systems and unprecedented as similar phenomena have not been realized in any other optical systems. The photon shuttling effect can be utilized to transfer optical information between multiple cavities for optical signal processing. The concept of mechanically mediated optical coupling can be extended to many other types of mechanical motions such as flexural plate waves and surface acoustic waves[36] to couple multiple photonic cavities over even longer ranges and in more sophisticated ways. Furthermore, increasing the mechanical Q of the current device to reduce the threshold pump level needed to start self-oscillation meanwhile reducing the inter-cavity coupling rate ($\kappa$) can potentially achieve single photon shuttling per mechanical cycle so the device can find application in quantum photonics. As the first multi-cavity optomechanical system the current device ($\Omega_\mathrm{m}/\gamma \simeq 2.3\times 10^{-5}$) is still far from the side-band resolved regime which is necessary to achieve backaction cooling of the mechanical resonator to quantum ground state and to observe other phenomenal photon-phonon dynamics[32,37]. However, it is foreseeable that in other systems that has reached the side-band resolved regime, such as optomechanical crystals that may consist of multiple photonic cavities[27], shuttling of single photons and inter-cavity Rabi oscillation of photons mediated by phonons are within reach. In such a regime, a wealth of quantum optomechanical effects can be expected to emerge. Finally, the photon see-saw device demonstrated here has a very high detection sensitivity of the torque moment determined to be $9.7\times 10^{-21}$ N·m/Hz$^{1/2}$ (see SI) and can be utilized immediately as an all-optically transduced torsional magnetometer[38,39], accelerometer[40] and gyroscope.

**Acknowledgement**

We acknowledge the funding support provided by the Young Investigator Program (YIP) of AFOSR (Award No. FA9550-12-1-0338). Parts of this work were carried out in the University of Minnesota Nanofabrication Center which receives partial support from NSF through NNIN program, and the Characterization Facility which is a member of the NSF-funded Materials Research Facilities Network via the MRSEC program. H. Li acknowledges the support of Doctoral Dissertation Fellowship provided by the Graduate School of the University of Minnesota.

**Author Contributions**

M. L. conceived and supervised the research; H. L. and M. L. designed the experiments; H. L. performed the fabrication and measurement; H. L. analyzed the data; M. L. and H. L. co-wrote the paper.

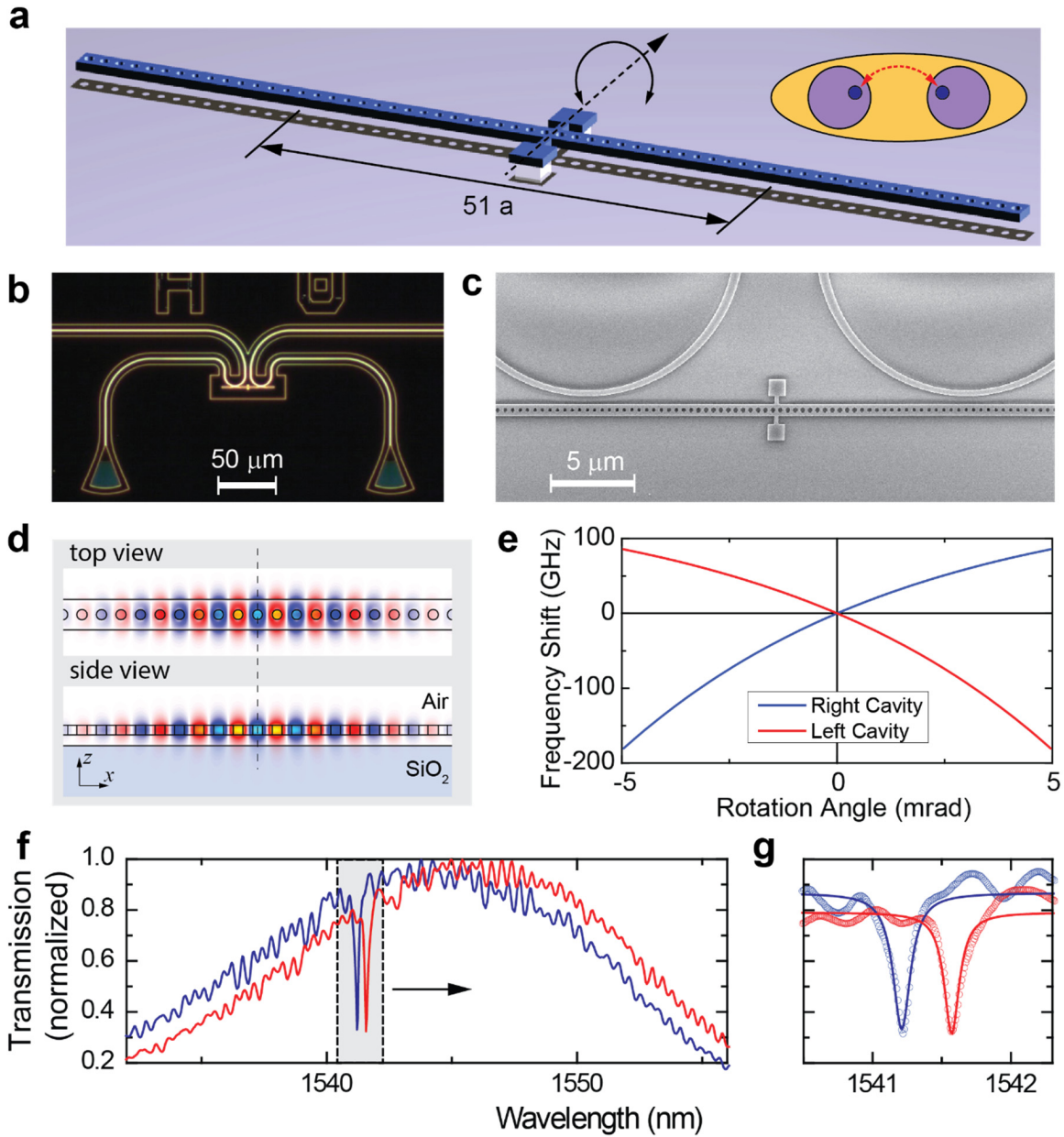

**Figure 1 Photon see-saw oscillator. a.** Artistic illustration of the photon see-saw oscillator consisting of a suspended nanobeam inscribed with two photonic crystal cavities, one on each side. $a$ = 450 nm is the lattice constant of the photonic crystal. For better visualization, this illustration is not to the scale of the actual device which consists of 102 air holes in total. The labeled distance between the centers of both sides of the nanobeam is consistent with the actual device, but not with the illustration. Inset: The torsional mechanical mode (yellow) extends to and couples two cavity optical modes (purple). **b.** Optical microscope image of the device, taken after the nanobeam was released, showing the integrated grating couplers connected to the coupling waveguides. **c.**

Scanning electron microscope image of the device, taken before the nanobeam was released, showing the two waveguides coupled with the two cavities separately. **d**. Finite-difference time-domain (FDTD) simulation generated field amplitude plot (top and side views) of the air-mode of the nanobeam photonic crystal cavity. The side view shows the mode field penetrating into the substrate to induce strong optomechanical coupling. **e.** Simulated relative frequency shift of the left and right cavities versus the rotation angle of the see-saw. The angular optomechanical coupling coefficients of the two cavities have opposite signs. **f.** and **g.** Broad (f) and narrow (g) band transmission spectra of the two cavities. The loaded optical quality factors are ~$1.0 \times 10^4$.

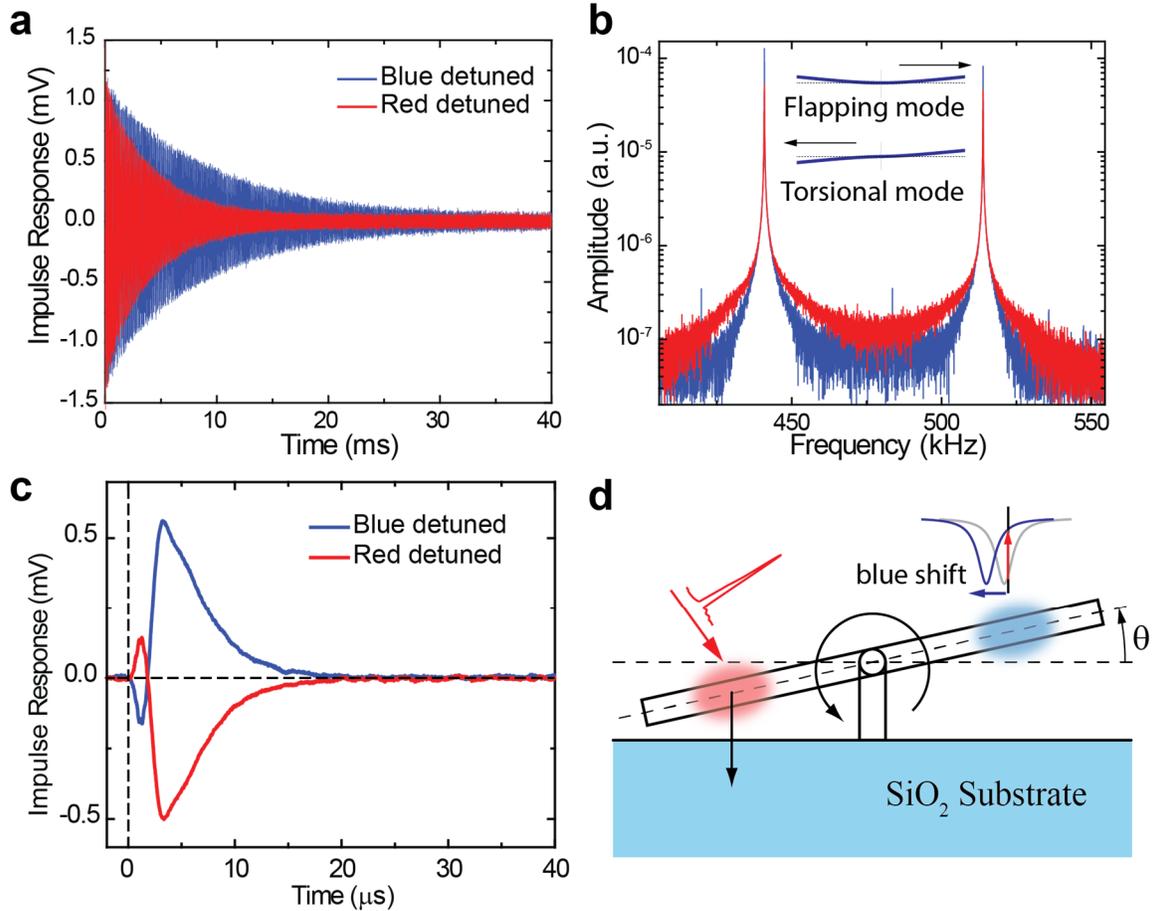

**Figure 2 Impulse response and photon see-saw effect. a.** Impulse response measured in vacuum when a pump pulse was sent to the left cavity and the transmission of a probe laser coupled to the right cavity is monitored. The response shows fast oscillation with a ring-down time which is noticeably longer when the probe is blue-detuned than when red-detuned, indicating the backaction cooling and amplification effects. **b.** Fourier transform spectra of the impulse responses, showing two out-of-plane mechanical modes: the torsional mode at 441 kHz and the flapping mode at 514 kHz. **c.** The impulse response measured at atmospheric pressure when fast mechanical oscillation is damped by air. The initial response (between 0 and 1.9 μs) indicates that the nanobeam rotated counter-clockwise under the impulse of the pump pulse and the resonance of the right cavity was blue shifted, as depicted in **d**. The later response (after 1.9 μs) is dominated by the thermo-optical effect.

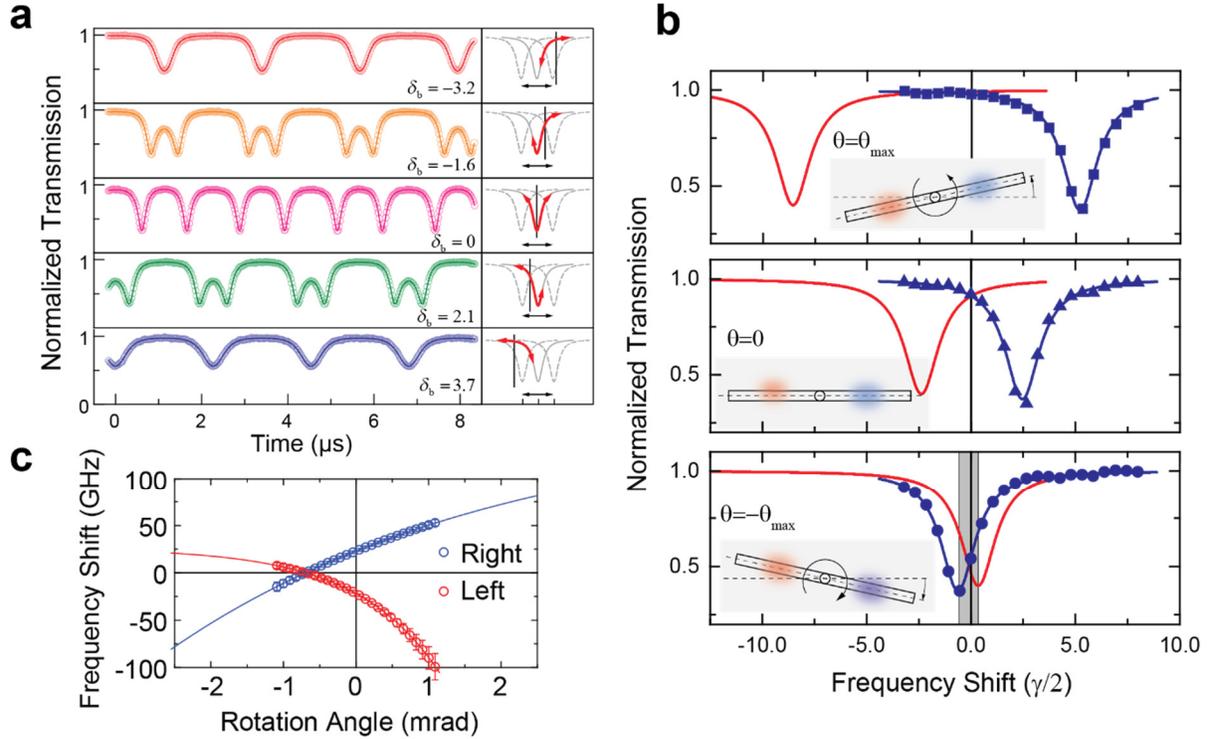

**Figure 3 Optomechanical self-oscillation and dynamics of the cavity resonances. a.** Time-domain traces of the probe laser transmission (normalized) from the right cavity with varying detuning ($\delta_b$) when optomechanical oscillation is excited by the pump laser coupled to the left cavity at a power level of 3.4 µW. The right column depicts the dynamics of the right cavity resonance (grey solid and dashed lines) and its alignment with the probe laser wavelength (black line). **b.** From the time-domain traces, stroboscopic snapshots of the cavity resonances can be constructed for the selected moments during the oscillation cycle. Shown are the positions of the resonances when the rotation angle $\theta = \theta_{max}$ (top), 0 (middle), $-\theta_{max}$ (bottom). It can be observed that at this amplitude, the resonances of the two cavities cross over (shadowed range) during the oscillation. **c.** The resonance frequencies of the two cavities versus the rotation angle of the nanobeam as determined from above results. The imperfect anti-symmetry of the behaviors of the two cavities is attributed to the non-ideal fabrication.

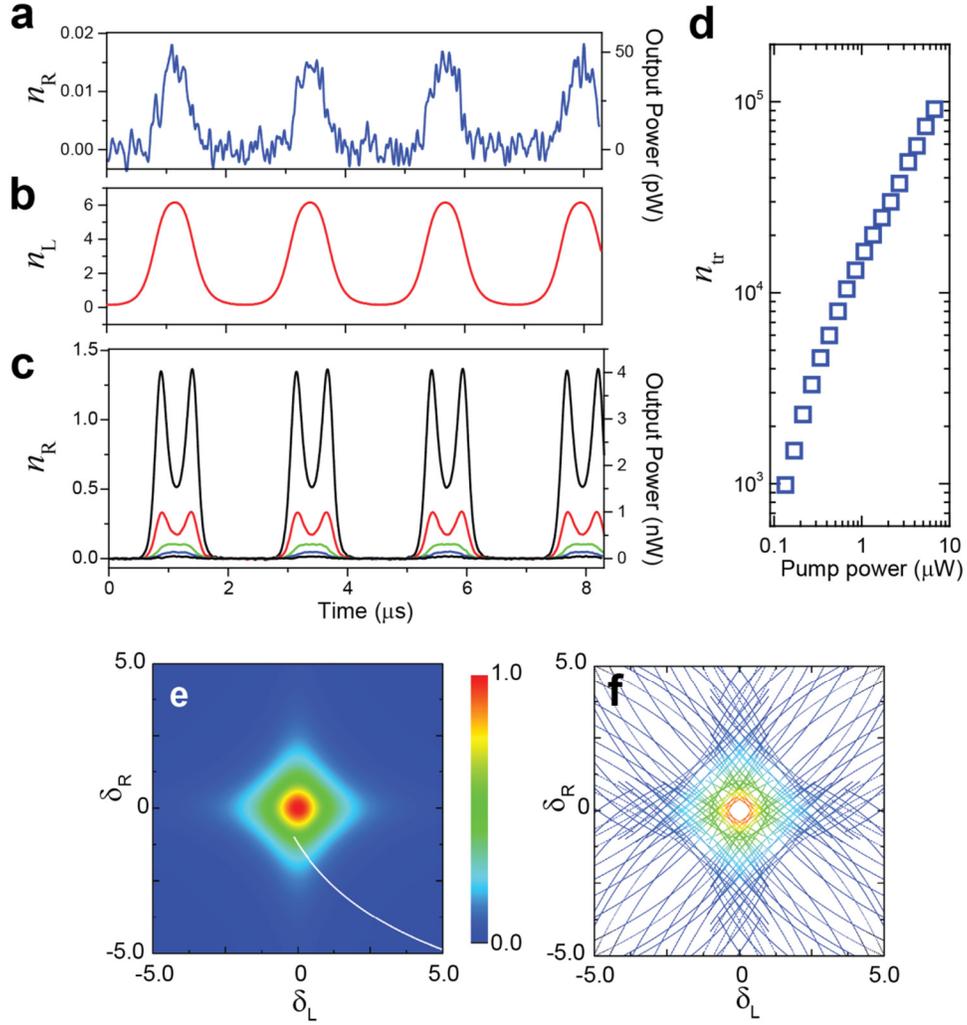

**Figure 4 Optomechanical photon shutting between two cavities. a.** and **b.** Time-domain traces of intra-cavity photon number of the right (**a**) and left (**b**) cavities when the pump laser power is at the threshold level of 0.135 µW. Also shown in the right axis of **a** is the output power of the right cavity. **c.** Intra-cavity photon number in the right cavity when the pump laser power increases from 0.135 to 6.76 µW. The right axis is the output power of the right cavity. **d.** Total number of photons shuttled from the left to the right cavity during one oscillation cycle versus pump laser power. The minimum is ~1000 photons per cycle when the pump is at the threshold level. **e.** and **f.** Theoretical (**e**) and experimental (**f**) plots of the right cavity's intra-cavity photon number $n_R$ versus the normalized pump laser detuning relative to the left ($\delta_L$) and the right ($\delta_R$) cavities. The white line in **e** represents the trajectory that the system undergoes when it is oscillating at the threshold level (i.e. the trace in **a**).